\documentstyle[12pt]{article}

\begin{document}
\newtheorem{algorithm}{PDP Algorithm}
\newtheorem{ralgorithm}{Relativistic PDP Algorithm}
\newcommand{\be}{\begin{equation}}
\newcommand{\ee}{\end{equation}}
\def\lqq{\lq\lq}
\def\rqq{\rq\rq}

\begin{titlepage}
%\today          \hfill
\begin{center}
%\hfill    Preprint No \\

\vskip .2in

{\large \bf Relativistic Quantum Events.}
%\footnote{Thanks text}
\vskip .50in

%alternate footnote for faculty:
%\footnote{}

\vskip .2in

Ph.~Blanchard${}^\flat$ \ and\ A.~Jadczyk${}^\sharp$\footnote{
e-mail: ajad@ift.uni.wroc.pl}

{\em ${}^\flat$ Faculty of Physics and BiBoS,
University of Bielefeld\\
Universit\"atstr. 25,
D-33615 Bielefeld\\
${}^\sharp$ Institute of Theoretical Physics,
University of Wroc{\l}aw\\
Pl. Maxa Borna 9,
PL-50 204 Wroc{\l}aw}
\end{center}

\vskip .2in

\begin{abstract}
Standard Quantum Theory is inadequate to explain the mechanisms
by which potential becomes actual.  Is inadequate and therefore
unable to describe generation of events.  Niels Bohr emphasized
long ago that the classical part of the world is necessary.
John Bell stressed the same point: that ``measurement"
cannot  even be defined within the Standard Quantum Theory,
and he sought a solution within hidden variable theories
and his concept of ``beables."

Today it is customary to try
to explain emergence of the classical world through a
decoherence mechanism due to ``environment".  But, we believe,
as it was with the concept of measurement, ``environment" itself
cannot be defined within the Standard Quantum Theory.

We have proposed a semi-phenomenological solution to this
problem by introducing explicitly, from the very beginning,
classical degrees of freedom, and by coupling these
degrees of freedom, through a Lindblad type coupling, to
the quantum world.  The resulting theory, we call
``Event Enhanced Quantum Theory". EEQT allows us
to describe an event generating mechanism for individual
quantum systems under continuous observation.  The objections
of John Bell are met and precise definitions of
an ``experiment" and of a ``measurement" have been given within EEQT.
However EEQT is, essentially, a non--relativistic theory.

In the present paper we extend the ideas of L.P. Horwitz
and C. Piron and we propose a relativistic version of EEQT,
with an event generating algorithm for spin one-half
particle detectors.  The algorithm is based on proper
time formulation of the relativistic quantum theory.
Although we use indefinite metric,  all the probabilities
controlling the random process of the detector clicks
are non--negative.
\end{abstract}
\end{titlepage}
\newpage
\section{Introduction.}

Quantum Theory has significantly changed our perspective on what Reality truly
{\em is}. Prior to Bohr and Heisenberg, speculations about Reality were in
the domain
of philosophy and not of physics. 
For a physicist, it was clear that there was a
Reality ``out there", and that physics was about making as precise a description
of this reality as possible.  This Reality had two kinds of concepts
connected to
it: static concepts and dynamic concepts.  Static concepts were
``objects," ``properties" of these objects, and ``relations" between the
objects.  Dynamic concepts were those of ``events"; defined as ``changes
of property," or ``changes of relation." And, ``change" was understood 
as ``change	in time".

With the advent of Quantum Theory the existence of this kind of simple reality
was first questioned and later denied.  Even if quantum physicists still
used the
concept of an ``object," it was no longer possible to speak about the actual
properties and relations of said objects.  A fortiori ``events" also disappeared
from the vocabulary of quantum theory.  Instead of objects and events, another
concept emerged as the dominant idea - that of ``measurement" or
``observation".

There were however, physicists who were not dominated by the new idea.
Some of them, the most prominent ones in this camp being
Albert Einstein and David Bohm, believed that quantum theory is a
temporary statistical description of some complex nonlinear
substructure - yet to be identified.  But opposition also appeared
in Quantum Theory's own camp.

John A. Wheeler stressed repeatedly \cite{wheeler}:
``No elementary quantum phenomenon
is a phenomenon until it is a registered (``observed," ``indelibly
recorded") phenomenon."  But, he did not give a definition of ``being
recorded" - and we now understand why? Because such a definition could
not be given within the orthodox quantum theory.

John Bell \cite{bell89,bell90} was the first to clearly realize that the
concepts
of ``measurement" and ``observation" were being used in Quantum Theory 
to brainwash
physicists into believing that these concepts are a part of Quantum
Theory, while in reality they belong to a metastructure.  He had
the courage to say \cite{bell87}: ``Either the wave function is not
everything or it is not right..." and he opted for an extended
theory that would contain, in addition to the quantum wave function,
classical variables, which he termed ``beables" - for ``being able to
be" \cite{bell-be}.

These theories were, however, inconsistent.  The classical part was acted
upon (by
the wave function) but there was no back reaction.  There was thus no way to
falsify these theories as no new results were predicted that would go beyond
those
predicted by the textbook recipes of the standard quantum theory.

It is our opinion that ``events" are classical in nature.  They obey the
classical ``yes-no" decision logic.  Even if the decisions are based
on fuzzy criteria - they are always sharp ``yes-no" decisions.  They are
points at which choices are being made.  Those who adhere to the Many
World Interpretation \cite{dewitt} would say: they are points at which
the Universe splits into branches.  Until we understand the true
nature of quantum theory and the true nature of time, these choices
must be considered as irreversible.  What was done in the past cannot be
undone later on.

In a series of papers (see \cite{blaja95a} and references there) we have
developed a semi--phenomenological theory of events - Event Enhanced Quantum
Theory (EEQT).  The theory is based on the linear time of Galileo and Newton. It
is a theory of a special kind of irreversible coupling between quantum and
classical systems.\footnote{Necessity of such a coupling
was envisaged already in the works by H. Primas \cite{prim1,prim2}
and A. Amman \cite{amman1,amman2}.}  Events are defined in this theory as
changes of
state of the classical subsystem.  They are accompanied by quantum
jumps - discrete changes of the quantum state.  In this theory, the
Schr\"odinger equation is replaced by a piecewise deterministic
algorithm that simulates the behavior of an individual quantum system
coupled to a classical ``measuring device".

We believe that the fundamental laws of Nature, that is the laws that Nature
herself uses while producing the events of the world ``out there," are based on
algorithms rather than on differential equations.  These algorithms are discrete
and probabilistic. The latter property mirrors the fact that we have
to describe an infinitely complex universe with the finite means that are, at
present, at our disposal.

A brief description of EEQT will be given in the the following Section.  The
simplest model that can be taken as an example for building more sophisticated
ones is the cloud chamber model developed in \cite{jad94b,jad94c}.  It describes
an irreversible coupling between a nonrelativistic quantum particle and an
arbitrary number of detectors sensitive to the presence of the particle in
localized areas of space.\\

From the very moment EEQT was born, we were aware of the unavoidable
difficulties
that might arise when the nonrelativistic theory would have to be replaced
by one
that is in agreement with Einstein's Relativity.  The difficulties to be
addressed in any attempt at building a relativistic theory of quantum
measurement have long been known by the experts. (see \cite{aha}, and
for a recent discussion \cite{peres95}). In a recent paper
\cite{blaja95f} we anticipated that the solution to this problem
would have to involve algorithms that are non-local not only in space but
also in time: ``However, if you try to work out a relativistic cloud
chamber model (...) the events must be also smeared out in the
coordinate time. (...). Nevertheless they can still be sharp in a
different ``time", called ``proper time" after Fock and Schwinger." The
proper time model of the cloud chamber is presented in Section 3.
We have chosen relativistic spin ${1\over2}$ particle, as its
proper--time quantum dynamics seems to cause more problems than
spin $0$ case - this due to the lack of Lorentz invariant
positive definite scalar product in the spin space. Our
solution, presented in Section 3, consists of using positive
detector coupling operators. In Section 2 we will recall the nonrelativistic
detector model: its standard version, as discussed in Refs.
\cite{blaja95a,jad94b,jad94c,blaja95e}, and also its  ``proper
time" version, so that a transition to the relativistic case
is made easier.

\section{Nonrelativistic Quantum Events.}

We reiterate, together with John A. Wheeler: No elementary quantum phenomenon is
a phenomenon until it is a registered phenomenon.  And, physics is about
phenomena.  The goal of physics is to understand Nature's phenomena; to be
able to
construct an ``artificial Nature" much in the way the goal of biology is to be
able to construct artificial life, the same way the ultimate goal of the
computer sciences is to create artificial intelligence.

But to construct artificial Nature we must be able to reproduce, in terms of
mathematical symbols, the natural phenomena.  In order to do this we first
must be
able to define within mathematics what a phenomenon is.  Or, more precisely,
what
a ``registered phenomenon" is.  For this purpose, we will call an elementary
registered phenomenon an {\em event}.  And now we must define what
constitutes an
event in terms of the mathematical structure.  We believe that it is
obviously impossible to do this within the standard mathematical framework of
the orthodox quantum theory.  The framework must be extended, the theory must be
enhanced.  And, it so happens that the extension is, in fact, only a slight one.
Significantly, nothing new is needed that has not already been discussed in the
framework of a general algebraic quantum theory.  The extension we have in mind
consists of allowing algebraic quantities that are not in any kind of
uncertainty
relation to any other quantity, that is, allowing {\em classical} (in
algebraic terms
they are called ``central") quantities. In other words: a phenomenon can be
registered, and an event can happen {\em only} if there is a classical subsystem
of the given quantum system.

It could be said, strictly speaking, that in a pure quantum world nothing could
or would ever ``happen."  But we see that things do happen in the world out
there.
This creates an obvious contradiction.  One possibility to get us out of the
contradiction is to negate events and phenomena, to call them ``illusion,"  to
admit only ``approximate events".  This is the road that most quantum physicists
of our day are prepared to take.

The other possibility is to accept that events do happen and to then introduce
explicitly a classical subsystem.  In this way,  an event is defined as a change
of state of this subsystem.  This then gives us  two further options.  One
option
consists of admitting only mental events. That is to say, events do exist, they
do happen, they are not illusions, but they happen only {\em in mind}.  This is
the option advocated by H. P. Stapp \cite{stapp93,stapp95}.

We are taking a less adventurous position by leaving the question 
``what is truly
and intrinsically classical?" open, or ``to be investigated."  Therefore our
theory, at the present stage, is semi-phenomenological.  That is, we are
building
models that aim at reflecting {\em some} of the mechanisms of Nature, but which
always can be improved by including more and more details in the description.

A general scheme of EEQT has been described in detail in Ref. \cite{blaja95a}
(see also \cite{blaja95e} for a short version). Here we will specify the scheme
to the particular but important case of a particle position detector.  We must
note that in the literature it is common to meet the opinion that any quantum
measurement can be, in the last instance, reduced to position measurements.
Although we do not share this conclusion, we believe that modelling particle
position detectors is important and allows us to understand the mechanism
through
which other types of quantum phenomena are being registered as events.

\subsection{Particle detectors}
We consider a nonrelativistic particle on a line. When no detector
is switched on then the quantum mechanical wave function $\Psi(x,t)$
representing the quantum state of the particle obeys the Schr\"odinger
equation:
\be
i\hbar {{\partial\Psi(x,t)}\over{dt}}= -{{\hbar^2}\over{2m}}
{{d^2\Psi(x,t)}\over {dx^2}} +
V(x,t)\Psi(x,t)
\ee
with a real potential $V(x,t)$.  In particular the norm of the wave function is
conserved.  Now, according to our philosophy, the Schr\"odinger equation
describes
a continuous evolution of ``possibilities".  Nothing happens.  No phenomena.
Now, let us add a particle detector that is coupled to the particle and can
``click" when the particle is ``nearby".  How to describe the situation of
the coupled pair particle+detector?  The detector itself is idealized as a
``yes-no" device.  As such it is a two-state classical system.  For simplicity
we will assume that the detector is at rest with respect to the
coordinate system we are working in.  When a quantum
particle is coupled to a classical detector then we are dealing with a hybrid
system.  Its pure states are described by pairs $(\Psi,\alpha)$, where
$\alpha=0,1$ describes the state of the detector.  Statistical states
of the total system are described by pairs $(\rho_0, \rho_1)$, where
$\rho_\alpha$ is a positive operator in $L^2(R)$ such that $Tr (\rho_0)+
Tr(\rho_1)=1$.  Now the coupling.  We will consider here a detector that
switches off after the first click and is no longer coupled to the
particle.  Other, more general types of particle detectors are
described in the references quoted in \cite{blaja95a}.  The coupling
is described by a positive operator $g$.  For a particle detector that is
sensitive to the particle position only, we will take for $g$ a function
$g(x)\geq 0$
which describes its spatial sensitivity.  One can think of $g(x)$ as a bell
like function localized
at the detector position.  Thus for a point like detector the support of $g(x)$
would shrink to a single point.  According to the general formalism of
EEQT described briefly in Section 1.1 of Ref. \cite{blaja95a}\footnote{Using
the notation of Ref. \cite{blaja95a} we put
$\alpha,\beta=0,1,$ $g_{10}=g, g_{01}=0, \Lambda_0=g^2, \Lambda_1=0$}
the Liouville equation describing the time evolution of the statistical
state of the total system reads:
\begin{eqnarray}
\left.
\begin{array}{ll}
{\dot \rho}_0&=-i[H,\rho_0]-{1\over2}\{\Lambda,\rho_0\},
\\
&\\
{\dot \rho}_1&=-i[H,\rho_1]+g\rho_0 g,
\end{array} \right\}
\end{eqnarray}
where $\Lambda=g^2$.
It follows from a general theorem proved in  \cite{jakol95}, that there is a
unique
Markov process on pure states of the total system that reproduces this
evolution of ensembles. It
is given by a piecewise deterministic algorithm (PDP of Ref.
\cite{blaja95a}) that
governs the click time of the counter.  In our case of a single counter the
algorithm reads:
\begin{algorithm}
Suppose that at time $t=0$ the
system is described by a (normalized) quantum state vector $\psi_0$ and the
counter
is off: $\alpha=0$.
Then choose a uniform random number $p_0\in [0,1]$, and proceed with
the continuous time evolution by solving the modified Schr\"odinger
equation
\be
{\dot \psi_t}=(-iH-{1\over2}\Lambda )\psi_t
\label{eq:tpsi}
\ee
with the initial wave function $\psi_0$ until $t=t_1$, where $t_1$
is determined by
$$\int_{0}^{t_1} (\psi_t,\Lambda \psi_t ) dt = p_0.$$ At $t=t_1$
the counter clicks, that is its state changes from $\alpha=0$ to
$\alpha=1$ and, at the same time, the state vector jumps:
$$\psi_{t_1}\rightarrow\psi_1=g\psi_{t_1}/
\Vert g\psi_{t_1}\Vert.$$
The evolution starts now again and it obeys the standard unitary
Schr\"odinger equation with the Hamiltonian $H$ - until the counter
starts its monitoring again, in which a case the continuous evolution
becomes again described by Eq.(\ref{eq:tpsi}), new random number $p_1$ is
selected, and so on.
\end{algorithm}

\subsection{``Proper time" formulation}
In this subsection we will give a four-dimensional formulation of
the nonrelativistic counter click model.  It will essentially be just
another formulation of the model above.  We will see that in a certain
limit it approximates the PDP algorithm of the previous subsection.

The Hilbert space we consider is now $L^2(R^2,dx dt)$ and the dynamics will
be given by a ``super-Hamiltonian"
\be {\cal H}=H-i{\partial\over{\partial}t}.
\ee
There is an extra time parameter associated with the counter, we will
denote it by $\tau$ and call it a ``proper time". The coupling between
the quantum particle and the counter is described by a positive operator
$G$ on our Hilbert space. The Liouville equation and the PDP algorithm
are much the same as above except that $x$ is replaced by $(x,t)$ and $t$
is replaced by $\tau$. \\
Let us now see how this formalism includes one in the previous
subsection.  With the notation as in Sec. 1.1 let us take for the
initial state $\Psi_0(x,t)$ a product:
\be
\Psi_0(x,t)=\phi(t)\psi_0(x),
\ee
where $\phi(t)$ and $\psi(t)$  are square integrable (with respect to $dt$
and $dx$ respectively) of norm one,
and $\Psi_0$ stands for $\Psi_{\tau=0}$.
Let us assume that $G$ depends on $x$ only:
\be
G(x,t)=g(x).
\ee
The equation (\ref{eq:tpsi}) for $\psi(x,t)$ is now replaced by
\be
{{\partial\Psi}\over{\partial\tau}}=(-iH-{\Lambda\over 2})\Psi
-{{\partial\Psi}\over{\partial t}},
\label{eq:ppsi}
\ee
which solves to:
\be
\Psi_\tau(x,t)=\phi(t-\tau ) e^{(-iH-{\Lambda\over2})\tau }\psi_0 (x).
\label{eq:ppsisol}
\ee
It is clear from this that identifying the coordinate time $t$ with the
``proper time" parameter $\tau$ we have the same inhomogeneous Poisson
process governing the detector click as in the previous subsection.

{\bf Remark} We have chosen the function $g$ to depend only on $x$ and
not on time for the reason that in this case the operator $\Lambda$
commutes with $\partial/\partial t$, and so Eq. (\ref{eq:ppsi})
is easily solved.

\section{Relativistic Quantum Events.}
Let us now consider the relativistic case.  Proper time formulation of
the relativistic quantum mechanics has been considered by many authors.  We
shall cite
here the classical paper by Horwitz and Piron \cite{horpir73}.  There
are also two review papers, one by Kyprianidis \cite{kyp87}, and one by
Fanchi \cite{fanchi93} (cf. also \cite{fanchi94}), where more references can
be found.
While the case of spinless particle was relatively straightforward, the case
of spin
$1/2$ caused interpretational problems because of the lack of Lorentz invariant,
positive definite, scalar product.  Several possible ways out of this
difficulty have
been considered.  Evans \cite{evans90} proposes to accept the indefinite scalar
product, while in \cite{horpireu75,pire78,reu79,arhor92} the authors introduce
an additional superselection (classical) variable ${\bf n}$ to parametrize a
family of
positive definite scalar product Hilbert spaces.  The latter was then discussed
by Horwitz and Arshansky \cite{hora82} who noticed that the Dirac operator
was not
Hermitian with respect to the positive definite products.  Our model will use
indefinite metric.  In fact, according to our philosophy, there is no reason at
all why scalar product is to be positive definite.  This is because we do not
start with the standard quantum mechanical probabilistic interpretation.  We
{\em derive} the interpretation from the coupling.  So, the only thing that
we have to worry about is that the probability of the detector click is to
be non-negative.  And we will see that
this is indeed the case.\\
We will take the standard representation of gamma matrices:
\be
\gamma^0=\pmatrix{I&0\cr0&-I\cr},\; \gamma^i=\pmatrix{0&\sigma^i\cr
-\sigma^i&0\cr}
\ee
and define an indefinite metric space by
\be
<\Psi , \Phi >=\int {\bar Psi}(x,t)\Phi(x,t) dx dt ,
\ee
where ${\bar\Psi}=\Psi^\dagger\gamma^0$ .
The Dirac matrices are Hermitian with respect to this scalar product, and so
is the
Dirac operator:
\be
{\cal D}= i\gamma^\mu (\partial_\mu+ieA_\mu)-m .
\ee
Let  us consider now a particle position detector which, for simplicity, is
at rest with respect to the
coordinate system. We associate with it the operator $G$ defined by
\be
(G\Psi)(x,t)={{I+\gamma_0}\over 2}g(x)\Psi(x,t),
\ee
where $g(x)$ is a positive, bell-like, function centered over the detector
position. \footnote{Notice that here,  as in the nonrelativistic case,
we assume that $g$ depends only on $x$ and not on $t$ - in the coordinate
system with respect the detector is at rest.}
It follows now that $G$ is positive, Hermitian with respect to
the indefinite metric scalar product, and the same holds for
$\Lambda=G^2$. We postulate the following relativistic version of the PDP
algorithm:

\begin{ralgorithm}
Suppose that at proper time  $\tau=0$ the
system is described by a quantum state vector $\Psi_0$ and the counter
is off: $\alpha=0$.
Then choose a uniform random number $p\in [0,1]$, and proceed with
the continuous time evolution by solving the modified evolution
equation

\be
{\dot \Psi}_\tau =(-i{{{\cal D}^2}\over{2M}}-{1\over2}\Lambda )\psi_t
\label{eq:taupsi}
\ee
with the initial wave function $\Psi_0$ until $\tau=\tau_1$, where $\tau_1$
is determined by
$$\int_{0}^{\tau_1} (\Psi_\tau ,\Lambda \Psi_\tau ) d\tau = p.$$ At
$\tau=\tau_1$
the counter clicks, that is its state changes from $\alpha=0$ to
$\alpha=1$ and, at the same time, the state vector jumps:
$$\Psi_{\tau_1}\rightarrow\Psi_1=G\Psi_{\tau_1}/
<\Psi_{\tau_1},G\Psi_{\tau_1}>.$$
If, after the first click, the detector is deactivated, then after the
click the evolution starts again and it obeys the standard
unitary Schr\"odinger equation with the Hamiltonian ${\cal H} = 
\frac{D^2}{2M}$.
\end{ralgorithm}
The algorithm contains second order, proper time, Dirac
equation.  This equation can be geometrically derived by the
method of dimensional reduction along an isotropic Killing
vector field in six dimensional space of signature $(++++,--)$ in
an exact analogy to the derivation of Levy-Leblond and Pauli
equation from five-dimensional space of signature $(++++,-)$ -
cf. Ref. \cite{jad94a}.
It is clear from the definition that the algorithm works well
and can be repeated.  Let us now assume that there are two detectors,
both at rest with respect to the coordinate system.  They are
both coupled to the quantum particle, the coupling operators
$G_i, i=1,2$ being given, as before, by:
\be
(G_i\Psi)(x,t)={{I+\gamma_0}\over 2}g_i(x)\Psi(x,t),
\ee
where the functions $g_i$ are localized at the detectors.
The operator $\Lambda$ is now a sum of two contributions:
\be
\Lambda=G_1^2+G_2^2.
\ee
The algorithm proceeds as before but now, when the event happens
at $\tau=\tau_1$ a decision must be made which of the two
detectors reacts.  The probability $p_i$ that the $i$-th
detector is activated is given by the same formula as in
the nonrelativistic case (cf. Ref. \cite{blaja95a} for a
general theory):
\be
p_i={{<\Psi_{\tau_1},G_i\Psi_{\tau_1}>}\over{
<\Psi_{\tau_1},G_1\Psi_{\tau_1}>+<\Psi_{\tau_1},G_2\Psi_{\tau_1}>}}.
\ee
Generalization to a larger number of detectors, that are not
necessarily at rest, is straightforward.
\section{Final Remarks.}
We have proposed a relativistic PDP algorithm that allows
one to model the behavior of a detector coupled to a relativistic
spin ${1\over2}$ particle.  Adding more detectors does not cause
any difficulties.  Our algorithm is repeatable, thus allowing for
a continuous monitoring of the particle position, as in the
nonrelativistic cloud chamber model described in \cite{jad94b,jad94c}.
It would be interesting to see how our relativistic event
generating algorithm can be used to test the idea of interference
in time as discussed by Arshansky, Horwitz and Lavie in \cite{arhola83}.

In the nonrelativistic case there is a dual description: by
a continuous in time Liouville equation that describes
time evolution of statistical states, and by
the PDP Algorithm that simulates Nature's event generation
for individual systems.  In the relativistic case we have
restricted ourselves to the individual description.  In fact,
at present, we do not know what would be the right mathematical
formalism and its physical interpretation for a relativistic
analogue of the Liouville equation.  Formally we can write an
equation as in the nonrelativistic case, but now using an
indefinite metric space.  Some of the relevant mathematics
have been developed in the past by one of us \cite{ajad70}.
What is yet to be done is to study the nonrelativistic
limit of the Relativistic PDP and to see that Nonrelativistic
PDP of EEQT is recovered this way.  The ideas given in the papers
by Horwitz \cite{hor91}
and Horwitz and Rotbart \cite{horrot81} can be applied
for studying such a limit.  Other problems that are yet
to be investigated are: quantum field theoretical
generalization along the lines of Ref. \cite{jad94c},
and an explicit formula for time--of--click probability
for a pointlike detector and a special intitial
wave packet - as in Ref. \cite{blaja95f}.  It is to be
observed that a quantum field theoretical version
will have to involve, as it was for one particle,
indefinite metric.

It must be noted that the Relativistic PDP, in cases of
more than one detector, involves {\em non-local} decision
making algorithm.  Thus, even if the detectors are treated as
classical in our approach, nevertheless there is no local
explanation for their behavior.  To decide which of the two
detectors will click at the event time involves a random
choice based on probabilities that are computed non-locally.
How Nature herself is doing this - is a big puzzle.  A solution
to this puzzle must be postponed till a later time when the
very nature and origin of the Planck constant, of space and of
time are better understood.

\section*{Acknowledgment(s)}
One of us (A.J.) would like to thank the A. von Humboldt Foundation for
support.  He would also like to thank L. Horwitz for encouragement
and discussion, and L. Knight for reading the manuscript.

\end{document}